\documentclass[conference]{IEEEtran}
\IEEEoverridecommandlockouts

\usepackage{cite}
\usepackage{amsmath,amssymb,amsfonts}
\usepackage{algorithmic}
\usepackage{algorithm}
\usepackage{bm}
\usepackage{braket}
\usepackage{graphicx}
\usepackage{textcomp}
\usepackage{xcolor}
\usepackage{multirow}
\usepackage{siunitx}
\def\BibTeX{{\rm B\kern-.05em{\sc i\kern-.025em b}\kern-.08em
    T\kern-.1667em\lower.7ex\hbox{E}\kern-.125emX}}

\newcommand{\crt}[1]{\hat{a}^\dagger_{#1}}
\newcommand{\dst}[1]{\hat{a}^{\phantom{\dagger}}_{#1}}
\newcommand{\bts}[1]{{\bf{#1}}}

\newcommand{\hartree}{E_{\mathrm{h}}}

\begin{document}

\title{ 

Closed-loop calculations of electronic structure on a quantum processor and a classical supercomputer at full scale

}

\author{\IEEEauthorblockN{Tomonori Shirakawa}
\IEEEauthorblockA{\textit{Center for Computational Science} \\
\textit{RIKEN}\\
Kobe, Japan \\
0000-0001-7923-216X\\
t-shirakawa@riken.jp}
\and
\IEEEauthorblockN{Javier Robledo-Moreno}
\IEEEauthorblockA{\textit{IBM Quantum} \\
\textit{IBM T.J. Watson Research Center}\\
Yorktown Heights, NY, USA \\
j.robledomoreno@ibm.com}
\and
\IEEEauthorblockN{Toshinari Itoko}
\IEEEauthorblockA{\textit{IBM Quantum} \\
\textit{IBM Research -- Tokyo}\\
Tokyo, Japan \\
0009-0006-2611-2301\\
itoko@jp.ibm.com}
\and
\IEEEauthorblockN{Vinay Tripathi}
\IEEEauthorblockA{\textit{IBM Quantum} \\
\textit{IBM T.J. Watson Research Center}\\
Yorktown Heights, NY, USA \\
0000-0003-1418-9046\\
tripathi@ibm.com}
\and
\IEEEauthorblockN{Kento Ueda}
\IEEEauthorblockA{\textit{IBM Quantum} \\
\textit{IBM Research -- Tokyo}\\
Tokyo, Japan \\
Kento.Ueda@ibm.com}
\and
\IEEEauthorblockN{Yukio Kawashima}
\IEEEauthorblockA{\textit{IBM Quantum} \\
\textit{IBM Research -- Tokyo}\\
Tokyo, Japan \\
0000-0001-5918-1211\\
yukio.kawashima@ibm.com}
\and
\IEEEauthorblockN{Lukas Broers}
\IEEEauthorblockA{\textit{Center for Computational Science} \\
\textit{RIKEN}\\
Kobe, Japan \\
lukas.broers@riken.jp}
\and
\IEEEauthorblockN{William Kirby}
\IEEEauthorblockA{\textit{IBM Quantum} \\
\textit{IBM T.J. Watson Research Center}\\
Yorktown Heights, NY, USA \\
William.Kirby@ibm.com}
\and
\IEEEauthorblockN{Himadri Pathak}
\IEEEauthorblockA{\textit{Interdisciplinary Theoretical and } \\
\textit{Mathematical Sciences RIKEN}\\
Wako, Japan \\
0000-0002-5919-8002\\
himadri.pathak@riken.jp}
\and
\IEEEauthorblockN{Hanhee Paik}
\IEEEauthorblockA{\textit{IBM Quantum} \\
\textit{IBM Research -- Tokyo}\\
Tokyo, Japan \\
hanhee.paik@us.ibm.com}
\and
\IEEEauthorblockN{Miwako Tsuji}
\IEEEauthorblockA{\textit{Center for Computational Science } \\
\textit{RIKEN}\\
Kobe, Japan\\
miwako.tsuji@riken.jp}
\and
\IEEEauthorblockN{Yuetsu Kodama}
\IEEEauthorblockA{\textit{Center for Computational Science} \\
\textit{RIKEN}\\
Kobe, Japan \\
yuetsu.kodama@riken.jp}
\and
\IEEEauthorblockN{Mitsuhisa Sato}
\IEEEauthorblockA{\textit{Center for Computational Science}\\
\textit{RIKEN}\\
Kobe, Japan \\
msato@riken.jp}
\and
\IEEEauthorblockN{Constantinos Evangelinos}
\IEEEauthorblockA{\textit{IBM Quantum} \\
\textit{IBM T.J. Watson Research Center}\\
Yorktown Heights, NY, USA \\
0000-0003-1418-9046\\
cevange@us.ibm.com}
\and
\IEEEauthorblockN{Seetharami Seelam}
\IEEEauthorblockA{\textit{IBM Quantum} \\
\textit{IBM T.J. Watson Research Center}\\
Yorktown Heights, NY, USA \\
sseelam@us.ibm.com}
\and
\IEEEauthorblockN{Robert Walkup}
\IEEEauthorblockA{\textit{IBM Quantum} \\
\textit{IBM T.J. Watson Research Center}\\
Yorktown Heights, NY, USA \\
walkup@us.ibm.com}
\and
\IEEEauthorblockN{Seiji Yunoki}
\IEEEauthorblockA{\textit{Pioneering Research Institute} \\
\textit{RIKEN}\\
Wako, Japan \\
yunoki@riken.jp}
\and
\IEEEauthorblockN{Mario Motta}
\IEEEauthorblockA{\textit{IBM Quantum} \\
\textit{IBM T.J. Watson Research Center}\\
Yorktown Heights, NY, USA \\
0000-0003-1647-9864\\
mario.motta@ibm.com}
\and
\IEEEauthorblockN{Petar Jurcevic}
\IEEEauthorblockA{\textit{IBM Quantum} \\
\textit{IBM T.J. Watson Research Center}\\
Yorktown Heights, NY, USA \\
0000-0003-1234-6386\\
petar.jurcevic@ibm.com}
\and
\IEEEauthorblockN{Hiroshi Horii}
\IEEEauthorblockA{\textit{IBM Quantum} \\
\textit{IBM Research -- Tokyo}\\
Tokyo, Japan \\
horii@jp.ibm.com}
\and
\IEEEauthorblockN{Antonio Mezzacapo}
\IEEEauthorblockA{\textit{IBM Quantum} \\
\textit{IBM T.J. Watson Research Center}\\
Yorktown Heights, NY, USA \\
mezzacapo@ibm.com}
}

\maketitle

\begin{abstract}
Quantum computers must operate in concert with classical computers to deliver on the promise of quantum advantage for practical problems. To achieve that, it is important to understand how quantum and classical computing  can interact together, and how one can characterize the scalability and efficiency of hybrid quantum-classical workflows. So far, early experiments with quantum-centric supercomputing workflows have been limited in scale and complexity. 
Here, we use a Heron quantum processor deployed on premises with the entire supercomputer Fugaku to perform the largest computation of electronic structure involving quantum and classical high-performance computing. We design a closed-loop workflow between the quantum processors and 152,064 classical nodes of Fugaku, to approximate the electronic structure of chemistry models beyond the reach of exact diagonalization, with accuracy comparable to some all-classical approximation methods. 
Our work pushes the limits of the integration of quantum and classical high-performance computing, showcasing computational resource orchestration at the largest scale possible for current classical supercomputers.
\end{abstract}

\section{Introduction and Background}
The electronic structure problem amounts to solving the static Schr\"{o}dinger equation ${\hat{H}|\Psi\rangle = E |\Psi\rangle}$. For molecular systems, it can be formulated on a discrete molecular basis set, where $\hat{H}$ is the Born-Oppenheimer Hamiltonian
\begin{equation}
\label{eq:es_ham1}
\hat{H} = \sum_{ \substack{pr\\\sigma} } h_{pr} \, \crt{p\sigma} \dst{r\sigma}
+ 
\sum_{ \substack{prqs\\\sigma\tau} }
\frac{(pr|qs)}{2} \, 
\crt{p\sigma}
\crt{q\tau}
\dst{s\tau}
\dst{r\sigma}
\,.
\end{equation}
Here we have defined the fermionic creation/annihilation operator $\crt{p\sigma}$/$\dst{p\sigma}$ associated to the $p$-th basis set element and the spin-$z$ polarization $\sigma$, while $h_{pr}$ and $(pr|qs)$ are the one- and two-body electronic integrals. The Hamiltonian is defined up to an additive constant.

For large systems that exhibit pronounced multireference character, solving the Hamiltonian Eq.(\ref{eq:es_ham1}) is a challenging task. While several classical approximation methods exist, like perturbative approaches, wavefunction ansatzes or Monte Carlo integration~\cite{leblanc2015solutions,motta2017towards}, all of them have a specific domain of applicability, depending on the method considered.

On the other hand, finding the low-energy spectrum of $\hat{H}$ is also a key application of quantum computing. The quantum phase estimation algorithm was initially proposed in the early days of quantum computing~\cite{kitaev1995quantum} to estimate the spectrum of many-body Hamiltonians such as Eq.~(\ref{eq:es_ham1}). However, it cannot scale on pre-fault-tolerant quantum processors due to large circuit size and substantial depth requirements. 

Instead of quantum phase estimation, many quantum algorithms have been proposed to enable the simulation of ground states in current pre-fault-tolerant quantum devices. Historically, the well-known Variational Quantum Eigensolver (VQE)~\cite{peruzzo2014variational} and its variants make use of the variational principle to construct an ansatz in the form of a parametric quantum circuit. However, for complex Hamiltonians such as the molecular problem, variational algorithms relying on the computation of expectation values on quantum devices face fundamental scaling limitations due to the high measurement overhead~\cite{wecker2015progress}. These issues have confined the scaling of quantum chemistry simulations to a few qubits~\cite{zhao2023orbital}. Recent developments in quantum-centric algorithms provide a pathway to scaling to larger systems with quantum sample-based workflows, such as quantum-selected configuration interaction~\cite{kanno2023QSCI}, which proposed to use quantum computers for selected-CI calculations, and the sample-based quantum diagonalization methods\cite{ robledo2024chemistry, yu2025quantum}, which led to reliable electronic structure simulations with quantum data for up to 85 qubits.

\begin{figure}[h!]
    \centering
    \includegraphics[width=0.7\linewidth]{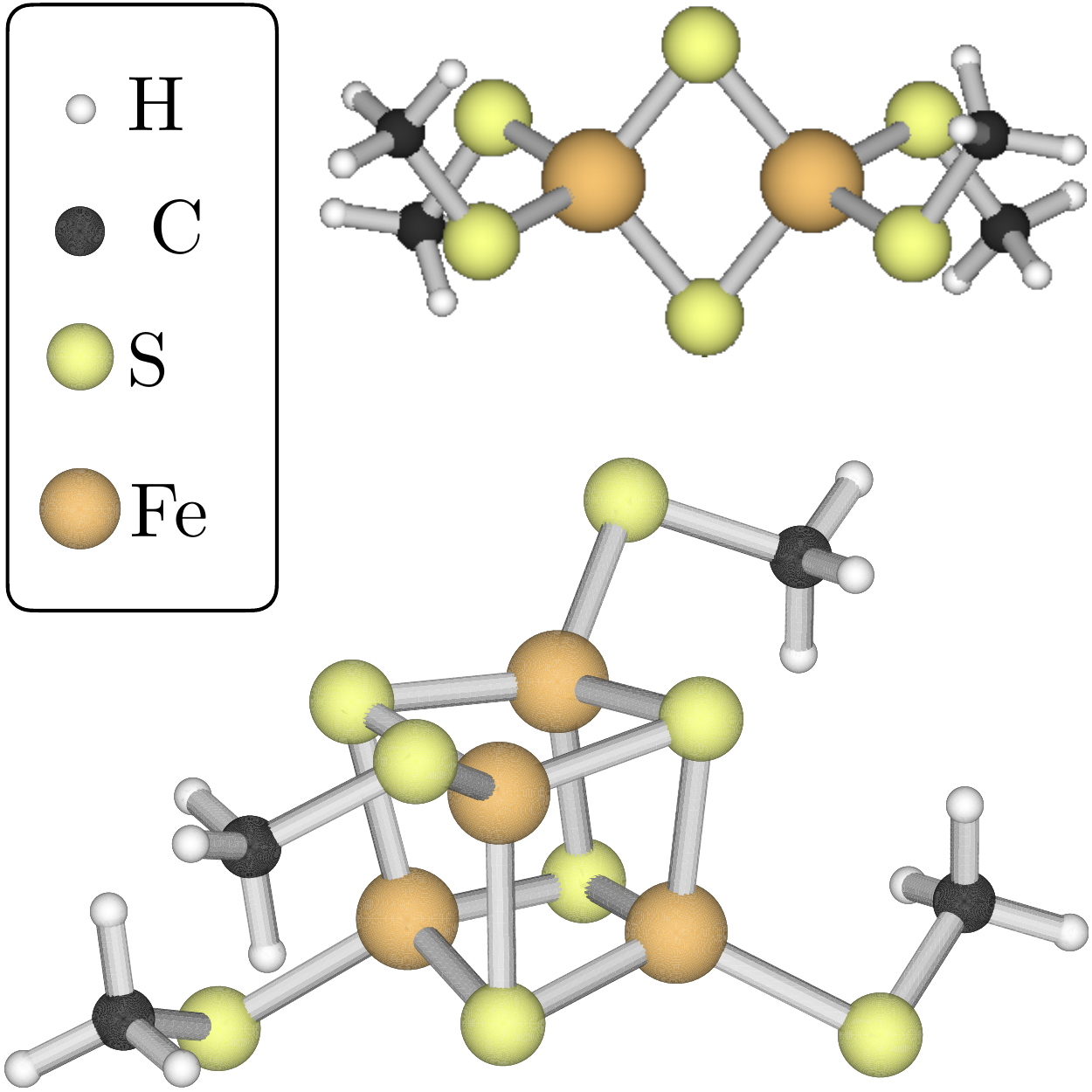}
    \caption{{\bf Geometries of the [2Fe-2S] (top) and [4Fe-4S] (bottom) clusters considered in this study.}}
    \label{fig:Fe4S4}
\end{figure}

Here, we target the ground-state properties of methyl-capped 
$\mathrm{[Fe_2S_2(SCH_3)_4]^{2-}}$ and 
$\mathrm{[Fe_4S_4(SCH_3)_4]^{2-}}$ iron-sulfur clusters -- henceforth abbreviated as [2Fe-2S] and [4Fe-4S] and whose geometries are depicted in Fig.~\ref{fig:Fe4S4} -- using active spaces of 50 electrons in 36 orbitals and 54 electrons in 36 orbitals, respectively.
Iron-sulfur clusters are a universal biological motif.
They are found in a variety of metalloproteins -- collectively referred to as iron-sulfur proteins -- which are characterized by the presence of one or more sulfide-linked di-, tri-, or tetra-iron centers in variable oxidation states, collectively forming a cluster. Iron-sulfur proteins include ferredoxins, hydrogenases, and nitrogenase, which participate in remarkable chemical reactions at ambient temperature and pressure~\cite{beinert1997iron, johnson2005structure}.


The electronic structure of iron-sulfur clusters -- featuring multiple low-energy wavefunctions with diverse magnetic and electronic character -- is key to their rich chemical activity. At the same time, it poses a formidable challenge for classical numerical methods.
Practical simulations -- Hartree-Fock and density functional theory -- rely on a mean-field approximation that treats only classical-like quantum states without entanglement. This approximation is fundamentally inadequate for iron-sulfur clusters, because the iron $3d$ shells are partially filled and near-degenerate on the Coulomb interaction scale, leading to strong electronic correlation in low-energy wavefunctions, and invalidating any independent-particle picture and the related concept of a mean-field electronic configuration. Broken-symmetry mean-field calculations -- the earliest approach used for these clusters -- only provide averaged properties across multiple electronic states and cannot resolve individual wavefunctions~\cite{sharma2014low}.

Accurate computation must instead involve entangled superpositions of multiple electronic configurations. In principle, the number of such configurations scales combinatorially with the number of iron and sulfur atoms in the cluster, and therefore quickly becomes intractable.
However, by employing methods that impose structure on these entangled superpositions~\cite{miralles1993specific,chan2011density}, it is possible to elevate quantum simulations from the mean-field level to the level of correlated many-body quantum mechanics required to describe the cluster's low-energy properties, while substantially mitigating the exponential complexity of a general quantum-mechanical formulation.

To investigate multireference ground-state wavefunctions in [2Fe-2S] and [4Fe-4S], we use active spaces of 50 electrons in 36 orbitals (of Fe[3d,4s,4p] and S[3p] character) and 54 electrons in 36 orbitals (of Fe[3d], S[3p], and ligand[$\sigma$] character). The dimensions of the corresponding Hilbert spaces, $\binom{36}{25}*\binom{36}{25} = 3.61 \cdot 10^{17}$ and $\binom{36}{27}*\binom{36}{27} = 8.86 \cdot 10^{15}$, are several orders of magnitude above the current scale of exact diagonalization, $1.0 \cdot 10^{12}$~\cite{gao2024distributed}.
As a result, meaningful electronic structure calculations must rely on approximate methods.
Furthermore, the breakdown of the mean-field approximation in iron-sulfur clusters implies that conventional methods for classical computers -- restricted Hartree-Fock (RHF), configuration interaction singles and doubles (CISD), and coupled cluster singles and doubles (CCSD) -- provide inaccurate results. Instead, the state of the art (SOTA) for classical electronic structure computations in iron-sulfur clusters is the density matrix renormalization group (DMRG)~\cite{sharma2014low} yielding ground-state energy estimates of $E_{\mathrm{DMRG,[2Fe-2S]}} = -5049.217~\hartree$ and $E_{\mathrm{DMRG,[4Fe-4S]}} = -327.239~\hartree$.

In the context of quantum computing, active spaces of electrons in 36 spatial orbitals require 72 qubits using the standard Jordan-Wigner map for fermion-to-qubit degrees of freedom~\cite{somma2002simulating}. A previous study~\cite{robledo2024chemistry} investigated the electronic structure of iron-sulfur clusters with the Sample-based Quantum Diagonalization (SQD) method in chemically realistic active spaces. This is the current SOTA for quantum computations on pre-fault-tolerant quantum computers, both in terms of problem size and accuracy: for the [4Fe-4S] cluster, it obtained an estimate for the ground-state energy of $-326.635~\hartree$, between RHF ($E_{\mathrm{RHF}} = -326.547~\hartree$) and above CISD ($E_{\mathrm{CISD}} = -326.742~\hartree$) -- two methods that, as discussed above, essentially break down in this situation. Understanding scaling of integrated quantum-classical methods is a mandatory requirement to match classical state-of-the-art methods in these and other, equally and even more challenging, instances of the electronic structure problem.

Our approach to approximate the ground states of [2Fe-2S] and [4Fe-4S] builds on the SQD method~\cite{robledo2024chemistry}, which uses quantum-centric supercomputing architectures~\cite{alexeev2023quantum}.
The active-space Hamiltonian Eq.~(\ref{eq:es_ham1}), for both [2Fe-2S] and [4Fe-4S], is mapped onto a 72-qubit operator using a Jordan-Wigner transformation.
Then, a quantum circuit prepares a wavefunction ansatz $|\Psi\rangle$, which approximates the support of the exact ground state. 
Recent experimentation of the SQD method for chemistry relied on the Local Unitary Cluster Jastrow (LUCJ) circuit~\cite{motta2023bridging}, with parameters derived from CCSD calculations.

\begin{figure*}[h!]
    \centering
    \includegraphics[width=.9\linewidth]{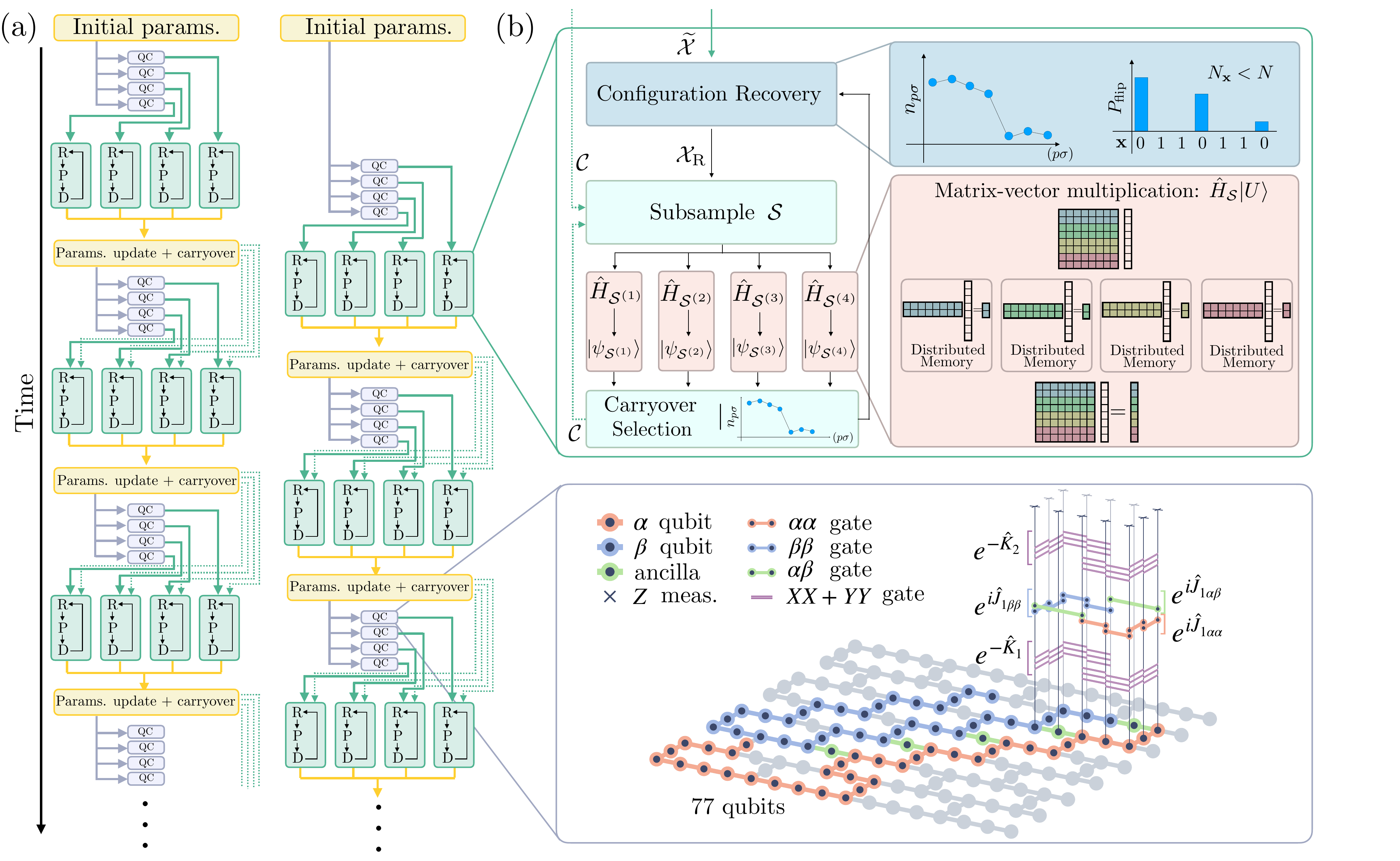}
    \caption{{\bf Depiction of the workflow used to optimize the quantum energies.} \textbf{(a)} High-level overview of the workflow to update the circuit parameters. The update of circuit parameters relies in a differential evolution algorithm with two populations of walkers, for a total of four walkers. The purple blocks labeled ``QC'' correspond to the execution of four quantum circuits associated to the walkers. The green blocks depict the SQD energy estimation for each of the walkers, which are executed in parallel in a HPC environment. Each of the green blocks executes three iterations of configuration recovery (R) with their corresponding projections (P) and diagonalizations (D). Purple arrows represent the loading of quantum circuits into the quantum processor. Green arrows represent the transfer of quantum measurement outcomes to the classical component. The dotted green lines denote the carryover of a fraction of configurations $\bts{x}$ with the largest SQD wave function component. \textbf{(b) Top:} The classical component of the workflow includes the configuration recovery step and projection and diagonalization in the subspace $\mathcal{S}^{(i)}$. The diagonalization step, achieved with a Davidson eigensolver. The red inset shows the distribution of the matrix-vector multiplication required by the Davidson eigensolver, where the application of each row of the matrix is distributed across a number of nodes of the supercomputer {\it Fugaku}. The massive parallelization allows to perform diagonalizations on $10^9$-dimensional subspaces in under 10 mins. \textbf{(b) Bottom:} Depiction of the LUCJ quantum circuit considered in this study, executed on {\it Heron} quantum processors, and the corresponding circuit layout.}
    \label{fig:workflow}
\end{figure*}

A set of quantum measurements $\mathcal{S}$ from the LUCJ quantum circuit first undergoes a configuration recovery step, and then is processed by a classical computer to project and diagonalize $\hat{H}$ on $\mathcal{S}$ 
\begin{equation}\label{eq:projection}
\hat{H}_{\mathcal{S}}=\hat{\mathcal{P}}_{\mathcal{S}}\hat{H}\hat{\mathcal{P}}_{\mathcal{S}}
\;,
\end{equation}
where the projector $\hat{\mathcal{P}}_{\mathcal{S}}$ is $
\hat{\mathcal{P}}_{\mathcal{S}} =\sum_{ {\bts{x}} \in \mathcal{S}} | {\bts{x}} \rangle \langle {\bts{x}} |.$ Under the Jordan-Wigner encoding, the sampled determinants are identified with the electronic configurations $\bts{x} \in \mathcal{S}$.
The size and constituents of $\mathcal{S}$ determine the accuracy of the approximate target eigenstate. The corresponding classical heuristics to search the set of relevant configurations to include in $\mathcal{S}$ fall under the general umbrella of Selected Configuration Interaction (SCI) methods~\cite{holmes2016heat, holmes2016efficient, smith2017cheap, sharma2017semistochastic}. Recent studies suggest that a quantum circuit $\Psi$ can produce accurate statistical models $p(\bts{x}) = \left|\langle \bts{x} | \Psi \rangle \right|^2$ to produce better quality subspaces than some SCI heuristics~\cite{robledo2024chemistry}. 

Existing pre-fault-tolerant quantum computers are subject to quantum noise.
In the present setting, the effect of the noise is to transform the ideal sample distribution $P_\Psi = \left|\langle \bts{x} | \Psi \rangle\right|^2$ to some other  $\widetilde{P}_\Psi$, the distribution from which we sample actual measurement results yielding a noisy set of configurations $\widetilde{\mathcal{X}}$.
This effective distribution introduces unwanted configurations that do not contribute to low-energy states. We can think of this as a broadening of the distribution $P_\Psi$, and consequently only some fraction of the sampled configurations $\widetilde{\mathcal{X}}$ correspond to actual quantum signal.
To mitigate this, a \emph{self-consistent configuration recovery} technique was introduced in~\cite{robledo2024chemistry} to probabilistically recover noiseless samples from the noisy measured set $\widetilde{\mathcal{X}}$, which is used to build $\mathcal{S}$.

The classical diagonalization step produces a wavefunction described by the expansion of a general many-body state in $\mathcal{S}$ (of polynomial size in the number of electrons and orbitals) of the basis of single-particle electronic configurations:
\begin{equation}
	\label{eq:Psi_sub}
    |\psi_\mathcal{S}\rangle = \sum_{\bts{x} \in \mathcal{S}} \psi_{\bts{x}} |\bts{x}\rangle,
\end{equation}
where, $\bts{x} \in \{0, 1\}^{2N_\textrm{orb}}$  are bitstrings whose length is twice the number of spatial orbitals $N_\textrm{orb}$. The first and second halves of each bitstring represent the occupancy of the spin-up and spin-down orbitals respectively: $|\bts{x}\rangle$ represents a Slater determinant -- or electronic configuration -- of the form $
|\bts{x}\rangle = \prod_{p,\sigma} \left(\hat{a}_{p\sigma}^\dagger \right)^{x_{p\sigma}} |0\rangle$, 
where $p = 1, 2, ..., N_\textrm{orb}$, and $\sigma \in\{\uparrow, \downarrow \}$. The coefficients of the expansion $\psi_{\bts{x}}$ are stored classically in a lookup table, and the energy of $|\psi_\mathcal{S}\rangle$ is used as a best approximation of the ground state energy.

\section{Methods}
Here we significantly improve on the SOTA quantum results of~\cite{robledo2024chemistry} with a combination of algorithmic innovations and optimal management of quantum and classical resources, improved quantum hardware, and large-scale deployment of classical computational resources. 

We implement a closed-loop optimization between quantum and classical processors. The differential evolution optimizer that we employ allows for an efficient orchestration of the quantum and classical computational resources used. This is achieved through a synchronization of the classical component of one population with the quantum execution of the other population, thus optimizing the computational resource utilization.

In order to achieve optimal resource utilization, the quantum and classical components must have comparable execution times. Since the quantum runtime is fixed by the repetition rate of the circuits ($10~\si{\micro\second}$), we require an efficient and parallelization-compatible projection and diagonalization implementation, whose runtime at fixed subspace dimension can be controlled by the number of parallel processes that are being executed. We achieve this by implementing a highly parallel Davidson diagonalization algorithm compatible with distributed memory environments, using up to 152,064 nodes of the supercomputer {\it Fugaku}.

\subsection*{Energy optimization workflow}\label{sec:workflow}

Empirically, the CCSD parameter initialization of the quantum wavefunction $|\Psi\rangle$ has been shown to yield accurate ground state energies in a variety of molecular systems in the 20-orbital scale, and only reasonable results once the 30-orbital mark is crossed~\cite{robledo2024chemistry, barison2025quantum, kaliakin2024accurate, liepuoniute2024quantum, shajan2024towards}. It is for these larger systems, like the iron-sulfur clusters that we consider here, where there is room for significant improvements. As the performance of SQD depends on parameters of the quantum circuit to sample, we optimize the parameters by a classical iterative algorithm, which determines the overall workflow. This algorithm must scale with parallel computing and also make it possible to simultaneously execute quantum and classical workloads, in order to utilize both computing resources as much as possible. We choose Differential Evolution (DE) \cite{storn1997differential}, which is an evolutionary algorithm, a type of metaheuristic well studied in the classical combinatorial optimization community. DE has a set of individuals, or walkers, who move around the parameter space to find better solutions. 

DE is a gradient-free optimizer which in our case requires cost function evaluations for the energy of $| \psi_\mathcal{S}\rangle$, see Eq.~(\ref{eq:Psi_sub}).
At each iteration, the cost function is evaluated on a number of walkers in parameter space.
We consider two populations, each consisting of four walkers. Each walker within a population collaborates to iteratively update their respective LUCJ parameters. We execute the cost function evaluations for the walkers in parallel, orchestrating the quantum and classical components of each population optimization in order to minimize idle times of computational resources, as depicted in Fig.~\ref{fig:workflow}.
For each walker, DE updates parameters by mixing with other walkers to create a new trial solution. There are several strategies for the update, among which we found the ‘best/2/bin’ strategy from \cite{back2023evolutionary} to work fairly well for our closed-loop optimization. 

Furthermore, we retain a certain number of determinants found in a given wavefunction $|\psi_\mathcal{S}\rangle$ over to the next iteration of the optimizer, ranked and selected according to the magnitude of their coefficients $\psi_\bts{x}$, upon reaching a total fraction of the previous wavefunction, named \emph{carryover ratio}. We find that this carryover helps with stability of the optimization with respect to statistical fluctuations and improving the energy.

The variational expressivity of the wavefunction ansatz in Eq.~\eqref{eq:Psi_sub}, can be improved by dressing $|\psi_\textrm{S}\rangle$ with parametric orbital rotations $|\psi_\mathcal{S}(\kappa)\rangle = U(\kappa) | \psi_\mathcal{S}\rangle$, with $U(\kappa) = \exp \left(\sum_{pr; \sigma} \kappa_{pr} \hat{a}^\dagger_{p\sigma} \hat{a}_{r\sigma} \right)$, a fermionic Gaussian unitary. The unitary character of $U(\kappa)$ is ensured by enforcing that the real-valued parameters that define the rotation are anti-Hermitian $\kappa_{pr} = -\kappa_{rp}$. The action of the Gaussian unitary on a determinant $|\bts{x}\rangle$ is to transform it to another determinant~\cite{Thouless1960}:
\begin{equation}
    U(\kappa) | \bts{x} \rangle = |\bts{x}' (\kappa)\rangle = \prod_{p\sigma} \left( \crt{p^\prime\sigma} \right)^{x'_{p\sigma}}|\bts{0}\rangle \,,\, \crt{p^\prime\sigma} = \sum_r \Phi_{rp}  \crt{r\sigma},
\end{equation}
with $\Phi$ the $N_\textrm{orb} \times N_\textrm{orb}$ matrix exponential of $\kappa$. The goal of the energy optimization workflow is to minimize the Rayleigh quotient:
\begin{equation}
   E(\mathcal{S}, \psi, \kappa) =  \frac{\langle \psi_\mathcal{S}(\kappa)|\hat{H}| \psi_\mathcal{S}(\kappa) \rangle}{\langle \psi_\mathcal{S}(\kappa)| \psi_\mathcal{S}(\kappa) \rangle }
\end{equation}
with respect to the subspace choice $\mathcal{S}$, the wave function coefficients $\psi_\bts{x}$ for all $\bts{x}$ in $\mathcal{S}$, and the parameters of the orbital rotation $\kappa$. For fixed $\mathcal{S}$ and $\kappa$, the optimal value for $\psi_\bts{x}$ is found in closed form by the projection and diagonalization of $\hat{H}^\prime(\kappa) = U^\dagger(\kappa) \hat{H} U(\kappa)$. Since $U(\kappa)$ is a fermionic Gaussian unitary, its action on the Hamiltonian can be computed efficiently:
\begin{equation}
\label{eq:es_ham}
\hat{H}^\prime(\kappa) = \sum_{ \substack{pr\\\sigma} } h^\prime_{pr} \, \crt{p\sigma} \dst{r\sigma}
+ 
\sum_{ \substack{prqs\\\sigma\tau} }
\frac{(pr|qs)^\prime}{2} \, 
\crt{p\sigma}
\crt{q\tau}
\dst{s\tau}
\dst{r\sigma}
\,
\end{equation}
with 
\begin{equation}
\begin{split}
h^\prime_{pr} &= \sum_{ab} h_{ab} \Phi_{ap} \Phi_{br} \;, \\
(pr|qs)^\prime &= \sum_{abcd} (ac|bd) \Phi_{ap} \Phi_{cr} \Phi_{bq} \Phi_{ds} \;.
\end{split}
\end{equation}

For fixed $\mathcal{S}$ and $\psi_\bts{x}$ values, the parameters in $\kappa$ are optimized by the \textit{Limited-memory Broyden–Fletcher–Goldfarb–Shanno} (L-BFGS) algorithm~\cite{Broyden1970BFGS, Goldfarb1970BFGS, Fletcher1970BFGS, Shanno1970BFGS}. Gradients with respect to the matrix elements of $\kappa$ are computed via automatic differentiation as proposed in Ref.~\cite{RobledoMoreno2023basisrotations}. The maximum number of iterations to update $\kappa$ is set to 1000, but we often observe convergence early in the iterative process. We remark that several first- and second-order methods~\cite{head1988optimization,roos1980complete,werner1985second,kreplin2019second,kreplin2020mcscf,smith2017cheap,sun2017general} exist to optimize the values of $\kappa$ in the context of multiconfigurational, complete active-space, and selected configuration-interaction self-consistent field calculations.

\subsection*{Execution of quantum circuits on {\it Heron} quantum processors}

Current pre-fault-tolerant quantum devices are prone to errors, which skew the ideal, noiseless sampling to a noisy distribution. As a direct consequence samples from quantum hardware contain determinants that are faulty and irrelevant to the diagonalization. We deploy various strategies to reduce the impact of noise on the quantum sampled distribution:
\begin{itemize}
\item In superconducting qubit architectures no two qubits, and hence gates between qubits, are identical due to  uncertainties in the fabrication process, RTE and cryo hardware, and calibration procedures. A major task in reducing noise is to select the best available qubits given circuit graph and the connectivity constraints of the device. We initialize LUCJ circuits that are hardware efficient for a heavy-hex connectivity and optimize the circuit layout using a modified version of mapomatic \cite{mapomatic}. Mapomatic uses the VF2++ search order heuristic to find all isomorphic graphs within the device connectivity map. Additionally, we update the backend noise model by measuring the two-qubit gate errors using layer-fidelity \cite{layerfidelity} and the state preparation and measurement (SPAM) errors through measurements of the all-zero and the all-one states across the device. These characterizations are performed as closely in time as possible to the actual quantum sampling experiments to ensure that the error model faithfully represents the device behavior. Single-qubit gate error rates are obtained from the backend-reported values measured via randomized benchmarking. These error rates are then used to build a cost-function for each isomorphic circuit graph:
\begin{equation}
    \mathcal{F}_\mathrm{circuit} = 1-\prod_{q,k} \varepsilon^{\mathrm{SQ}}_{q,k}\prod_{g,k}\varepsilon_{g,k}^{\mathrm{TQ}}\prod_m\varepsilon_m^{\mathrm{SPAM}}
\end{equation}

where $\varepsilon^{\mathrm{SQ}}_{q,k}$, $\varepsilon^{\mathrm{TQ}}_{g,k}$ and $\varepsilon_m^{\mathrm{SPAM}}$ are the single-qubit errors, two-qubit errors and SPAM errors, respectively, as they appear for each operation within a given circuit. Finally, we map the LUCJ circuit onto the qubit subset that yields the highest estimated circuit fidelity, $\mathcal{F}_\mathrm{circuit}$. For every subsequent LUCJ parameter update (walker/population), the qubit layout remains fixed. 

\item LUCJ circuits have significant idle periods due to the structure of the circuits and the real-time duration of gates. Qubits, despite sitting idle, still interact with noise channels from the environment and the qubit state decoheres adding to the reduction in the sampling fidelity. Dynamical decoupling (DD)~\cite{Viola1999} is a technique that can be used to suppress decoherence and some of the qubit-qubit crosstalk errors. We use a sequence of four pulses $XX\bar{X}\bar{X}$ (super CPMG) in all the idling gaps to improve the fidelity of our sampling experiments, where $X$ and $\bar{X}$ rotate by $\pi$ and $-\pi$ (respectively) about the x-axis of the Bloch sphere. Note that we implement $\bar{X}$ using the symmetric compilation in which an $X$ gate is applied between virtual $Z$ gates of $\pm \pi$ rotations~\cite{vezvaee2025}.

\item We use the configuration recovery routine as detailed in~\cite{robledo2024chemistry}, with two improvements. First, by providing the initial electron occupancy distribution externally with MP2 perturbation theory, we perform configuration recovery even in the first iteration. Additionally, while subsampling quantum measurements, we set the batch number to 1 and instead perform diagonalization using a larger number of configurations.

\end{itemize}

\subsection*{Classical diagonalization on the supercomputer {\it Fugaku}}

The most computationally demanding part of the classical computation is the computation of the lowest eigenpair of the Hamiltonian projected over the space of selected configurations. To achieve this goal we employ the Davidson's method~\cite{davidson1975iterative}, wherein a set of vectors is produced iteratively: starting from a set of $d$ vectors $S=\{ {\bf{v}}_k \}_{k=0}^{d-1}$, one computes the lowest-energy eigenpair $(E,| {\bf{w}} \rangle)$ in $S$ and the residual vector $| {\bf{r}} \rangle = H | {\bf{w}} \rangle - E | {\bf{w}} \rangle$, applies a preconditioner operator to $| {\bf{r}} \rangle$ (here, the standard diagonal preconditioner ~\cite{morgan1986generalizations}) and uses it to expand $S$. The algorithm starts from ${\bf{v}}_0$ defined as the sampled configuration with lowest energy, and terminates when $\| {\bf{r}} \|$ is below a threshold (signaling that an eigenpair has been found) or a maximum number of iterations is reached. The most expensive part of the Davidson's method is the application of $\hat{H}$ to a vector in $S$, which is a sparse matrix-vector multiplication.

To ensure this is done as efficiently as possible and in sync with the quantum runtime, we have developed a code designed for use on massively parallel supercomputing platforms. The primary advantages of leveraging such large-scale parallel architectures are twofold: the ability to handle wavefunction vectors whose size exceeds the memory capacity of a single node, and the significant acceleration of computation through parallelized Hamiltonian operations on these vectors.
We distribute a single wavefunction vector across multiple nodes and parallelize the Hamiltonian application over this distributed data structure. In the following, we describe the strategy used for the parallelization of the Hamiltonian operation.

Let $\psi_b = \langle {\bf x}_b \vert \psi \rangle$ denote the wavefunction vector after the Hamiltonian application, and $\phi_b = \langle {\bf x}_b \vert \phi \rangle$ denote the vector before the operation. The Hamiltonian operation can be expressed as
\begin{equation}
    \psi_b = \sum_{b'} H_{bb'} \phi_{b'}
\end{equation}
where $H_{bb^{\prime}} = \langle {\bf x}_b \vert \hat{H} \vert {\bf x}_{b^{\prime}} \rangle$. 
Since $\psi_b$ is distributed across multiple nodes, evaluating the above summation requires the transfer of $\phi_{b'}$ data from other nodes. By replicating $\psi_b$ across different node groups, the computation of the summation over $b'$ can be parallelized across these replicated domains. On the other hand, since $\psi_b$ is already distributed, the computation across row indices $b$ can naturally be parallelized, and the replication of $\psi_b$ across node groups further increases the granularity of this parallelism. We do all these operations at double precision.

To enable scalable Hamiltonian operations on distributed wavefunction vectors, we employ a flexible three-dimensional communicator split method. Our implementation introduces three types of MPI communicators: 
(1) a basis (column index) communicator that manages the distribution of the wavefunction vector  across a basis space of size $B$; (2) a task communicator that coordinates $T$ replicated processes to parallelize the summation over $b'$; 
and (3) a row communicator that handles further division into $M$ partitions along the row index $b$. 

Our computational basis is defined as the product space spanned by the $\alpha$- and $\beta$-determinant configurations. 
Accordingly, in the implementation, the basis communicator $B$ has a size corresponding to the product of the two determinant spaces, expressed as $B = B_\alpha B_\beta$, 
where $B_\alpha$ and $B_\beta$ represent the communicators associated with the $\alpha$ and $\beta$ determinants, respectively, and we have the option of partitioning each of those communicators separately.

This configuration allows the wavefunction vector to be distributed over $B \times T \times M$ nodes. 
Neglecting communication overhead, each decomposition direction supports near-ideal scaling with respect to the number of nodes. Among these, only the basis communicator requires non-blocking neighbor-wise data shifts, which introduce relatively low overhead, while the replicated communicators (task and row) require synchronizing blocking global reductions on very large vectors via \texttt{MPI\_Allreduce}. 
To improve load balancing, we also introduce a new task assignment method that dynamically allocates work based on the size of each task, ensuring more uniform workload distribution across all processes. This is essential to get high efficiency when executing workflows at full scale on {\it Fugaku}. 

For our diagonalization task, which is shared with configuration interaction (CI) chemistry workflows, it is essential to efficiently evaluate only the non-zero elements of the Hamiltonian, whose sparse structure is determined by the Slater–Condon rules. The wavefunction is represented by configurations encoding the occupation of alpha- and beta-spin orbitals, and matrix elements are non-zero only when the corresponding determinants differ by one- or two-particle excitations~\cite{knowles1984new}. While conventional selected CI codes accelerate this process by precomputing and storing allowed excitations in intermediate arrays~\cite{holmes2016heat,sharma2017semistochastic}, our implementation differs by a key distinction: the CI space is defined as the tensor product of subspaces spanned by the alpha and beta components. As a result, all entries in the intermediate arrays become non-zero, enabling a simpler and more efficient implementation.
 
Since our optimization algorithm proceeds sequentially with the aforementioned wavefunction carryover, the previously obtained ground state is used as the initial seed for the next iteration step. However, because the subspace changes between iterations, only the components in the common subspace are used to construct the initial vector.

\subsection*{Classical resources}
In total, the classical component of the workflow used up to 152,064 nodes of the supercomputer {\it Fugaku}, each with 48 cores and 32 GiB of memory for a total of 7.3 million cores and 4.6PiB of memory. The SQD energy of four walkers of the population is evaluated simultaneously, with each walker performing a diagonalization in a distinct subspace. 
These four diagonalizations run in parallel, each making use of up to $38,016$ nodes, with the only exception of calculations where the subspace dimension was $10^{10}$ calculation, where we used 2 walkers and 76,032 nodes per walker.
In this demonstration, we considered two populations of up to four walkers each for DE.

To update the hyperparameters of DE, we used a mutation scaling factor $F=0.25$ and crossover ratio $C\!R=0.7$.
For population~0, for one walker, we used the same initial LUCJ parameters from CCSD as proposed in~\cite{robledo2024chemistry} (CCSD initial parameters) and, for the other walkers, we used initial parameters computed from double excitation amplitudes with random perturbations of $\pm 0.05$ from the amplitudes obtained by CCSD.
For population~1, we used the CCSD initial parameters for all walkers.

For each iteration of the configuration recovery step, we used the occupancies from the previous step except for the first iteration where we used the initial occupancies. The initial occupancies was computed classically, using one-particle reduced density matrix from the second-order Møller–Plesset perturbation theory (MP2).

We enabled wavefunction carryover across all diagonalizations for each walker, between optimization iterations and between multiple diagonalizations for configuration recovery within a single SQD energy evaluation.
After obtaining a noisy set of configurations $\widetilde{\mathcal{X}}$ via quantum sampling, we applied configuration recovery and randomly selected $D$ configurations from the recovered samples.

Given the spin-symmetric nature of the [2Fe-2S] and [4Fe-4S] clusters, its configuration exhibits mirror symmetry between the first and second halves corresponding to $\alpha$- and $\beta$-spin orbitals, respectively. 
Accordingly, after subsampling, the $D$ configurations were split into two halves, and unique half-configurations were selected. If the Hartree–Fock configuration $\ket{1^{\otimes 27} 0^{\otimes 9}}$ was not among the half-configurations, it was manually added. The resulting $D_h$ half-configurations were used in diagonalization.
These were then used to form a $D_h \times D_h$ subspace, $[ {\bf x}_b^h ]_{b=1}^{D_h} \otimes [ {\bf x}_{b^{\prime}}^h ]_{b^{\prime}=1}^{D_h}$, on which we performed diagonalization to obtain the ground-state wavefunction $\psi_{bb'}$.
From this wavefunction, we selected $c D_h$ half-configurations $[\tilde{\mathbf{x}}^h_b]_{b=1}^{c D_h}$ for carryover with carryover ratio $c$, prioritizing those with the largest weights $r_b = \sum_{b'} |\psi_{bb'}|^2$.
In our numerical experiments, $D_h$ was set to 3,200; 12,000; 32,000; and 10,000, corresponding to subspace dimensions of $1.0 \times 10^7$, $1.4 \times 10^8$, $1.0 \times 10^9$, and $1.0 \times 10^{10}$ respectively.

For the diagonalization part, the maximum number of Davidson iterations was limited to 10 steps. The iteration process was terminated when either of the following two conditions was met: (1) the desired accuracy was reached within 10 steps, or (2) the predefined maximum wall-clock time was exceeded. The convergence criterion was based on the norm of the residual vector, with the iteration stopping when it dropped below $1.0 \times 10^{-3}$. While this threshold typically corresponds to an energy error of about $1.0 \times 10^{-6}$, our actual target is a looser energy accuracy on the order of $1.0 \times 10^{-3}$. Therefore, this residual norm threshold is in fact rather stringent for our purposes. In practice, even when the iteration was terminated due to the wall-clock time limit, the energy error at termination was still around $1.0 \times 10^{-6}$, confirming that the solution had already reached sufficient accuracy. Furthermore, since the ground state obtained in the previous iteration is used as the initial guess for the subsequent iteration, convergence is usually achieved in fewer than 10 steps.

The performance of the diagonalization also depends on the configuration of MPI communicators. In the experiment involving a subspace of size $10^9$ using 16,384 nodes, four MPI processes were assigned to each node, and the wavefunction vector was distributed over 64 nodes, so $B$=256. To accelerate the computation, parallelization over the row and column indices was employed, with the number of partitions set to $T$=8 and $M$=32, respectively. Consequently, for each of the four walkers, the diagonalization was carried out using 16,384 MPI processes across 4,096 nodes.
In the experiment with a larger subspace of size $10^{10}$, we used 152,064 nodes with two MPI processes per node and four walkers 
($B=24\times24$, $T=12$, $M=11$). 
For comparison, a subset of the runs was performed using 76,032 nodes with two walkers under the same communicator configuration.
A summary of the parameters used in all experiments, including those described above, is provided in Table~\ref{tab:comp_setup} for reference.
In our numerical experiments, we observed that increasing the subspace dimension by a factor of 10 tends to result in approximately a 20–25$\times$ increase in classical computational cost. The actual computation times for each setup, reflecting this scaling behavior, are summarized in the table.

\begin{table}[t]
  \centering
  \caption{Summary of classical computing setups used in each subspace diagonalization. The table lists the subspace dimension, communicator parameters ($B$, $T$, $M$), number of walkers, total nodes used, and the actual time required for the diagonalization.}
  \label{tab:comp_setup}
  \begin{tabular}{cccccc}
    \hline
    Subspace & ($B$,$T$,$M$) & Walkers & Total Nodes & Time [min] \\
    \hline
    $10^{7}$  & (16,2,3)     & 4 & 384   &  $\sim$ 2  \\
    $10^{8}$  & (36,6,6)   & 4 & 5,184 &  $\sim$ 5 \\
    $10^{9}$  & (64,8,32)  & 4 & 65,536 & $\sim$ 10  \\
    $10^{10}$  & (576,12,22)  & 4 & 152,064 & $\sim$ 15-20 \\
    $10^{10}$  & (576,12,22)  & 2 & 76,032 & $\sim$ 15-20 \\
    \hline
  \end{tabular}
\end{table}

\subsection*{Quantum resources}

\begin{figure}[h!]
    \centering
    \includegraphics[width=1\linewidth]{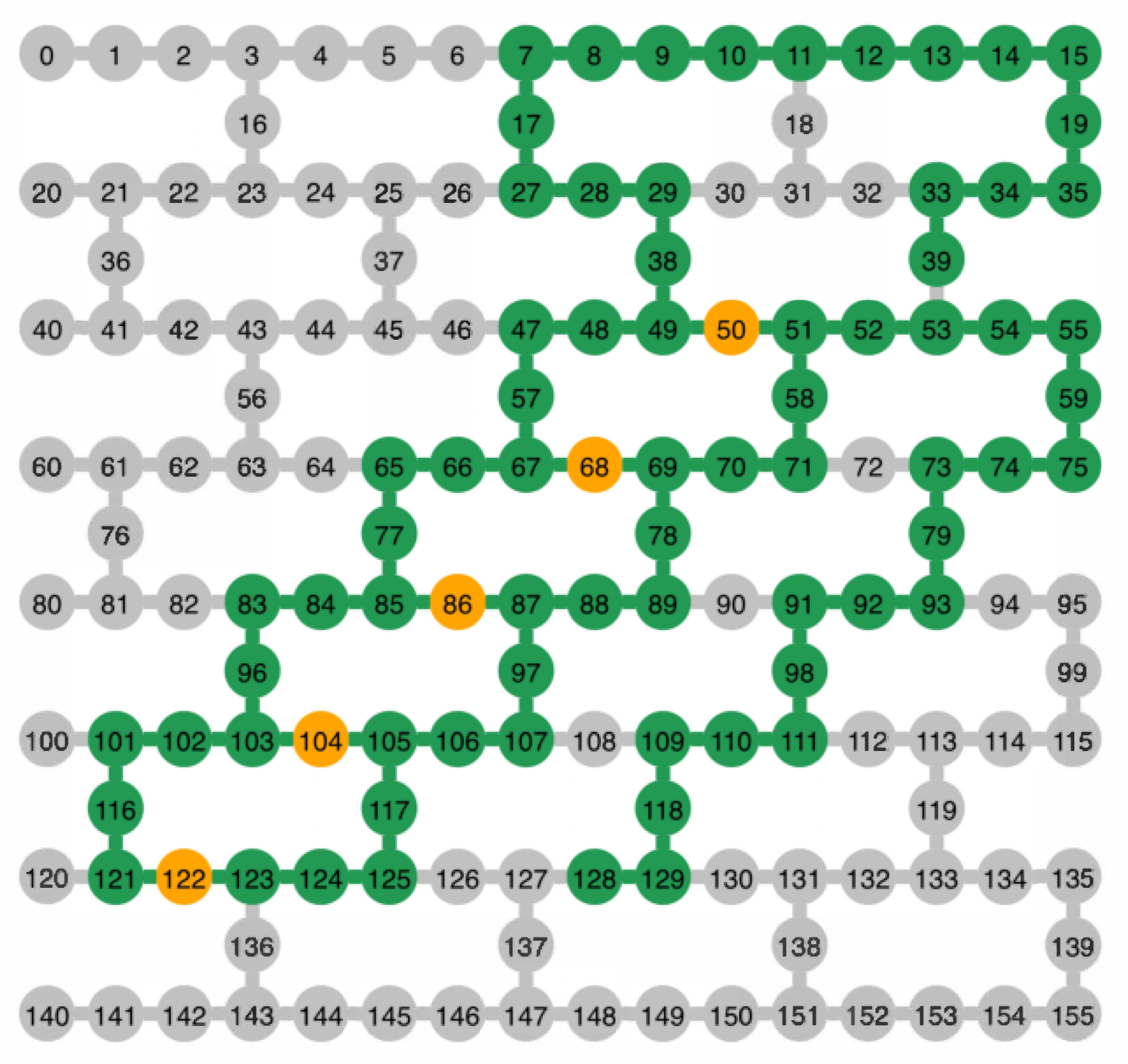}
    \caption{{\bf Schematic of the Kobe quantum processor.} Out of the full chip, 77 qubits are employed in the experiment, with 72 primary qubits highlighted in green and 5 ancillary qubits in orange; the remaining qubits are shown in gray.
    }
    \label{fig:device}
\end{figure}

The quantum experiments were executed on the \textit{ibm\_marrakesh} and \textit{ibm\_kobe}, two {\it Heron} quantum processors with $156$ superconducting qubits laid out in a heavy-hex connectivity. Fig.~\ref{fig:device} provides a comprehensive overview of the \textit{ibm\_kobe} architecture. The top panel illustrates the device connectivity. From the available $156$ qubits, we use $77$ qubits of which $2 \times 36 = 72$ are used to encode the LUCJ circuit and $5$ qubits are used as auxiliary qubits to mediate the density-density interactions between the two spin populations. $72$ primary qubits are marked in green and $5$ ancillary qubits in orange, while the qubits that do not take part in the circuit are depicted in gray. 

\subsection*{Implementation of the workflow}

Our approach maximizes the utilization of both quantum and classical computing resources while maintaining a simple workflow by overlapping their execution as shown in Algorithm~\ref{alg_de_impl}. The process involves two primary functions: \textsc{RunQuantum} and \textsc{RunClassical}.

\begin{figure}[!t]
\begin{algorithm}[H]
	\caption{Synchronous Quantum and Classical computations for Two Populations in Differential Evolution}
	\label{alg_de_impl}
\begin{algorithmic}[1]
    \STATE \textbf{async} \textsc{RunQuantum}($itreration$=0, $population$=0)
    \STATE \textbf{async} \textsc{RunQuantum}($itreration$=0, $population$=1)
    \FOR{$itr \gets 0$ \textbf{to} $MaxItr - 1$}
    \STATE \textbf{wait} \textsc{RunQuantum}($itreration$=$itr$, $population$=0)
    \STATE \textsc{RunClassical}($itreration$=$itr$, $population$=0)
    \IF{$itr + 1 < MaxItr$}
    \STATE \textsc{RunQuantum}($itreration$=$itr + 1$, $population$=0)
    \ENDIF
    \STATE \textbf{wait} \textsc{RunQuantum}($itreration$=$itr$, $population$=1)
    \STATE \textsc{RunClassical}($itreration$=$itr$, $population$=1)
    \IF{$itr + 1< MaxItr$}
    \STATE \textsc{RunQuantum}($itreration$=$itr + 1$, $population$=1)
    \ENDIF
    \ENDFOR
\end{algorithmic}
\end{algorithm}
\end{figure}

\textsc{RunQuantum} is invoked asynchronously for each population. It generates a LUCJ circuit based on parameters determined by a differential evolution algorithm and subsequently executes this circuit on a quantum processor.

Concurrently, while \textsc{RunQuantum} is processing a population on a quantum processor, \textsc{RunClassical} performs several classical tasks for a different population. These tasks include: recovering configurations from previous quantum computations, selecting relevant configurations, and performing a projected diagonalization. Crucially, \textsc{RunClassical} also drives the differential evolution process by generating new parameters that will be used by \textsc{RunQuantum} in its subsequent calls for the next iteration.

Our implementation (see our code~\cite{sqd-diag}) employs a hybrid programming approach to optimize performance and resource utilization. The quantum computation component, \textsc{RunQuantum}, is implemented in Python, while the classical processing component, \textsc{RunClassical}, is implemented in C++ for performance-critical operations. Table~\ref{tab:software} lists the exact software versions used in these experiments. Within \textsc{RunQuantum} (Python), we utilize Qiskit 1.4.2~\cite{aleksandrowicz2019qiskit} for constructing LUCJ circuits, transpiling these circuits for execution on specific quantum hardware, and retrieving the resulting measurements after quantum computation. We employ qiskit-ibm-runtime 0.37.0 to send quantum circuits and retrieve the output measurements. Additionally, we use ffsim 0.0.52 for the initial construction of the LUCJ circuits based on the provided parameters. For the classical computation component, \textsc{RunClassical} (C++), we have rewritten performance-critical parts of the configuration recovery and measurement subsampling functionality, originally found in the Qiskit addon of SQD 0.10.0, in C++ to enhance execution speed. The entire differential evolution workflow, including parameter generation for the LUCJ circuits used by \textsc{RunQuantum}, is also implemented in C++. The Python-based \textsc{RunQuantum} is integrated into the C++-based differential evolution workflow by invoking it through system calls, enabling the asynchronous execution of quantum computations managed by the classical C++ framework. This architecture allows us to benefit from the ease of use and rich quantum circuit manipulation capabilities of Qiskit and ffsim in Python. At the same time, we maximize the performance of computationally intensive classical tasks like configuration recovery, subsampling, diagonalization, and the core differential evolution algorithm through our custom C++ implementation.

\subsection*{Hybrid time performance measurement}
To effectively evaluate the utilization of both quantum and classical resources, we track the timing of each step within \textsc{RunQuantum} and \textsc{RunClassical}. Leveraging the metadata provided in the results of our experiments with Qiskit, we accurately determine the start and end times of quantum circuit execution on the hardware.

Within our analysis, we categorize the key steps of \textsc{RunClassical} as follows: \textbf{throw} encompasses the processes of constructing, optimizing (transpiling), and sending the quantum circuit to the quantum computer; \textbf{retrieve} involves serializing and receiving the results (measurements) from the quantum computer; \textbf{pre-processing} serializes the obtained bit strings, subsamples them, and recovers configurations for subsequent diagonalization; and \textbf{diagonalization} calculates the energy of the molecule based on the processed measurements. By timing these distinct stages in both the quantum (\textsc{RunQuantum}) and classical (\textsc{RunClassical}) computation, we gain insights into the efficiency of our hybrid approach and identify potential bottlenecks in resource utilization.

\section{Results}

\begin{figure*}[h!]
    \centering
    \includegraphics[width=\hsize]{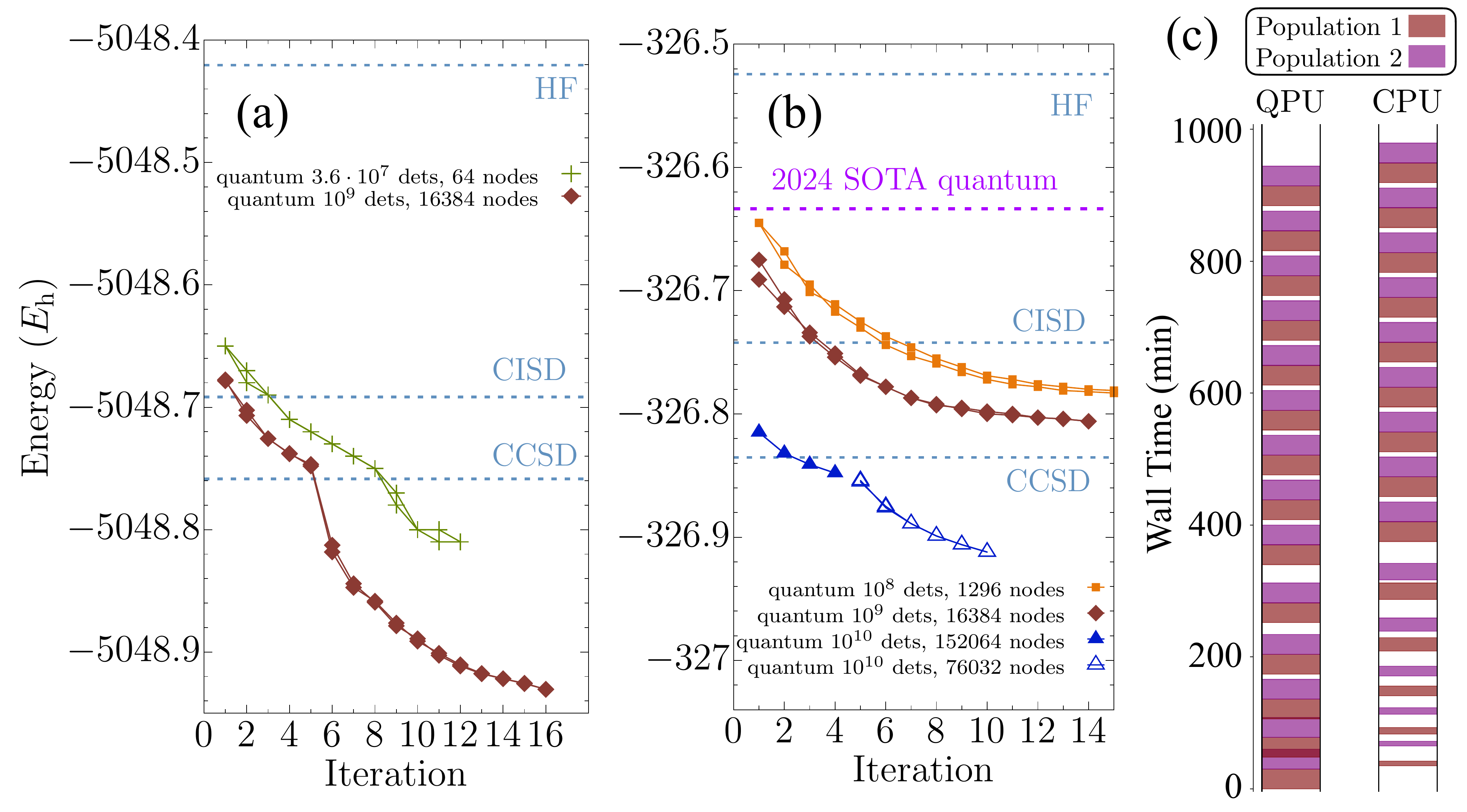}

    \caption{
    \textbf{{(a)}}
    {\bf Electronic structure calculations using Heron quantum processors and the supercomputer Fugaku.} Quantum energies (markers) as a function of the optimization step (a) for the [2Fe–2S] cluster, and (b) for the [4Fe–4S] cluster.
    The calculations for the [4Fe–4S] cluster with $10^8$ and $10^9$ determinants were performed on IBM Marrakesh, whereas the $10^{10}$ case and all [2Fe–2S] calculations were carried out on IBM Kobe.
    For reference, we also show the previous state-of-the-art (SOTA) quantum energy~\cite{robledo2024chemistry}, as well as the energies obtained from Hartree–Fock (HF), configuration interaction with singles and doubles (CISD), and coupled cluster with singles and doubles (CCSD).
    (c) Quantum ({\it Heron}  processor) and classical ({\it Fugaku} supercomputer) resource usage as a function of elapsed wall time. The orchestration of quantum and classical resources aims at minimizing idle (white) time. The data is from the $10^8$ determinants subspace from (b).}
    \label{fig:results}
\end{figure*}

Figure~\ref{fig:results}(a) and (b) present the results obtained by executing the closed-loop  workflow for a series of iteration steps.  Figure~\ref{fig:results}(a) corresponds to the [2Fe–2S] cluster, while Fig.~\ref{fig:results}(b) shows the results for the [4Fe–4S] cluster.  For reference, Fig.~\ref{fig:results}(b) also includes the previous state-of-the-art (SOTA) results obtained with quantum data~\cite{robledo2024chemistry}, and both panels (a) and (b) show the energies computed with Hartree–Fock (HF), configuration interaction with singles and doubles (CISD), and coupled cluster with singles and doubles (CCSD).
In both systems, the walker energies in each population decrease monotonically with iterations, eventually achieving lower energies than those obtained by CCSD.  This indicates that the quantum data can capture correlations beyond single and double excitations from the reference HF state.

For the [2Fe–2S] cluster, the lowest energy of $-5048.932~\hartree$ was obtained after 15 iterations using a subspace dimension of $10^9$, while for the [4Fe–4S] cluster, the lowest energy of $-326.912~\hartree$ was achieved after 10 iterations using a subspace dimension of $10^{10}$.

In the experiments on the [4Fe–4S] cluster using $10^8$ and $10^9$ determinants, we used four walkers for each of the two populations (0 and 1) and plotted the energy trajectories of each walker in population 1. For these runs, we repeated three configuration recovery and diagonalization steps within a single SQD iteration, and set the carryover ratio to 0.75.
In contrast, for the $10^{10}$ determinants run, each SQD iteration included only one configuration recovery and diagonalization step, and the carryover ratio was set to 0.5.
Using \textit{ibm\_marrakesh}, we acquired 750,000 measurement samples for each LUCJ circuit in the $10^8$ and $10^9$ runs. For the $10^{10}$ run, we used \textit{ibm\_kobe} and acquired 500,000 samples per circuit.
The $10^{10}$ determinants experiment began with a warm start using the LUCJ parameters and carryover bitstrings obtained at the final step of the $10^9$ run. We also incorporated the orbital optimization procedure described in Section~\ref{sec:workflow}.
In the first five iterations, the algorithm was executed independently for each population. In the latter five iterations, we introduced a cooperative scheme: at the end of each iteration, carryover bitstrings from one population were shared with the other. Within each population, the better-performing walker---selected from the two walkers---provided the energy and carryover bitstrings for the next iteration.
This cooperative strategy effectively doubled the number of diagonalizations and orbital optimizations per iteration. As a result, a noticeable difference in energy descent behavior was observed between the first and second halves of the run, as shown in the blue plots in Fig.~\ref{fig:results}(b), which represent energies for population 1.
Note that one walker per population was used during the first five iterations, whereas two walkers per population were used in the latter five iterations. The best point at $10^{10}$ determinants improves on the best existing results~\cite{robledo2024chemistry} by $278~m\hartree$.

For the experiment on the [2Fe–2S] cluster, we employed the same algorithm setup as in the latter five iterations of the [4Fe–4S] cluster experiment with $10^{10}$ determinants, except that the number of measurement samples per LUCJ circuit was reduced to 200,000.
In Fig.~\ref{fig:results}(a), we plotted the energy trajectories of two walkers from population 1. For this active space of [Fe2-S2], we also confirm that quantum data has a signal sufficient to obtain energies better than classical  CCSD methods for a subspace of $10^9$ determinants.

\begin{figure}
\centering
\includegraphics[width=\hsize]{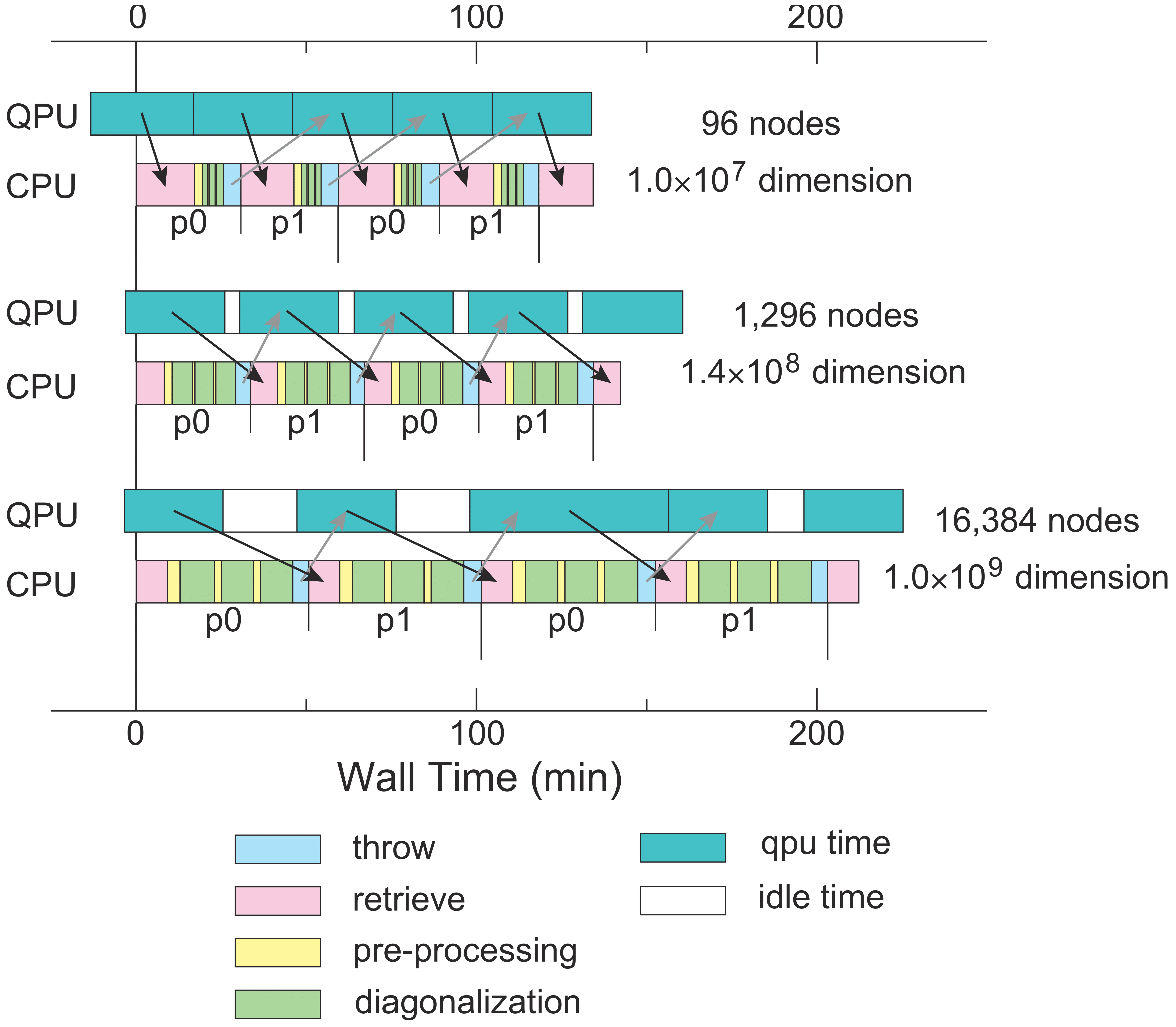}
\caption{{\bf Wall time profile of QPU and CPU usage.}
Computations for population 0 (p0) and population 1 (p1) are alternately executed on the classical processor. Here \textbf{throw} is the processing to prepare and send instructions to the quantum computer, \textbf{retrieve} receives output from the quantum computer; \textbf{pre-processing} prepares the list of configurations for diagonalization, and \textbf{diagonalization} performs projection and diagonalization to obtain molecular energies. 
}
\label{fig:time_profile}
\end{figure}

Figure~\ref{fig:results} (c) shows, as a function of the elapsed wall time, the quantum and classical resource utilization by the walkers of both populations.
As the iteration number increases, so does the classical wall time required to diagonalize the Hamiltonian projected into the subspaces proposed by the quantum circuits. Since the subspace dimension is kept fixed, this is a consequence of the increased number of off diagonal elements added to $\hat{H}_{\mathcal{S}}$, which indirectly signals the increased quality of the explored subspaces, resulting in better energies. After the first 5 iterations the idle time of quantum and classical resources is minimized, showcasing a nearly optimal synchronization in the scheduling of both quantum and classical resources.
This is further showcased in Figure~\ref{fig:time_profile}. We report the utilization of both quantum and classical resources by walkers in a typical iteration step, as a function of the number of classical nodes used and the dimension of the subspace. Since the quantum runtime is mostly fixed - except one execution where we observed a delay in the cloud access of the device -  by setting fixed number of quantum measurements per walker, independent of the subspace size, one can increase the number of classical nodes to keep the diagonalization time manageable even as the subspace dimension increases.

\begin{figure}[h!]
\centering
\includegraphics[width=\hsize]{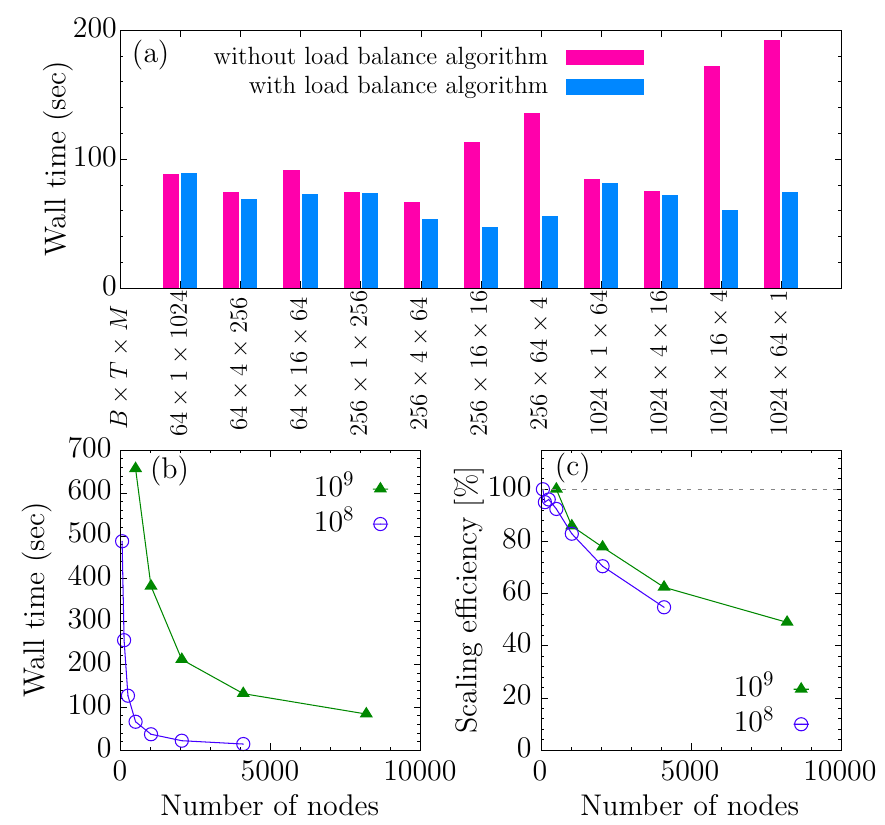}
\caption{ {\bf Classical parallelization performance.}
{\bf (a)}
Comparison of wall time-to-solution with and without a load balancing algorithm applied to the task communicator under different node assignments. The benchmark is based on a $10^{10}$-dimensional subspace derived from measurements obtained from a quantum computer. The total number of nodes is fixed at 16,384 (4 times larger than at $10^9$ for a single walker), with four MPI processes assigned per node. The communicator sizes $B$, $T$, and $M$ respectively correspond to the sizes of basis communicator, the task communicator for partitioning over column indices, and the row communicator for partitioning over row indices. We use this test to determine the parameters used in the experiments in Table~\ref{tab:comp_setup}.
{\bf (b)} Wall-clock time to solution and {\bf (c)} strong scaling performance of the diagonalization as functions of the number of nodes, for problems with subspace dimensions of $1.0 \times 10^9$ and $1.0 \times 10^8$, constructed based on measurements generated by a quantum computer.
}
\label{fig:diagtime}
\end{figure}

As mentioned in Section \ref{sec:workflow}, we introduce a flexible three-dimensional communicator decomposition and a load-balancing algorithm for task assignments over column indices in order to improve the parallel performance of the diagonalization step. To evaluate the impact of this strategy, we compared the wall-clock time required to reach a solution with and without the load-balancing algorithm under fixed node and MPI process configurations, as shown in Fig.~\ref{fig:diagtime} (a). Note that the total number of processes is fixed. Increasing $T$ (the communicator size along the column index) sometimes causes performance degradation in the absence of load balancing. On the other hand, by introducing the load-balancing algorithm, we observed a reduction in execution time of up to 16 processes for the task-parallel communicators.
Based on these results, in our actual experiments we chose the diagonalization parameters such that the communicator size $T$ was as close as possible to $\sqrt{B}$, where the best performance was observed. Although the available number of nodes sometimes limited us from realizing the exact configuration, we selected parameters that approximated this optimal condition as closely as possible.

To evaluate the parallel performance of the diagonalization, we also measure the wall-clock time required to obtain the solution for problems with subspace dimensions of $10^8$ and $10^9$, constructed based on measurements obtained from the quantum processor. The results are shown in Fig.~\ref{fig:diagtime} (b). As the number of nodes increases, the time required to reach a solution decreases accordingly. For example, when using 512 nodes, the wall time is 65.899 seconds for the $10^8$-dimensional case, while it is 656.574 seconds for the $10^9$-dimensional case.

To assess parallelization efficiency, we show strong scaling performance, based on wall-clock time, in Fig.~\ref{fig:diagtime} (c). In practical computations, the performance achieved at larger node counts was approximately 60–70\% of that at the baseline (smaller) node count. It is worth noting, however, that in quantum–classical hybrid computations, the quantum part is typically executed within a fixed time frame of several tens of minutes (e.g., ~30 minutes).
Therefore, the diagonalization process must be designed to complete within the timeframe dictated by the quantum execution.
This temporal constraint, arising from the need to synchronize with the quantum execution schedule, inherently limits the granularity of parallel task decomposition. Nevertheless, it reflects a deliberate design choice aimed at efficient coordination between quantum and classical resources. The observed scaling performance should thus be understood not as a limitation of the implementation itself, but rather as a consequence of balancing architectural synchronization with practical execution requirements.

\begin{figure}[h!]
\centering
\includegraphics[width=\hsize]{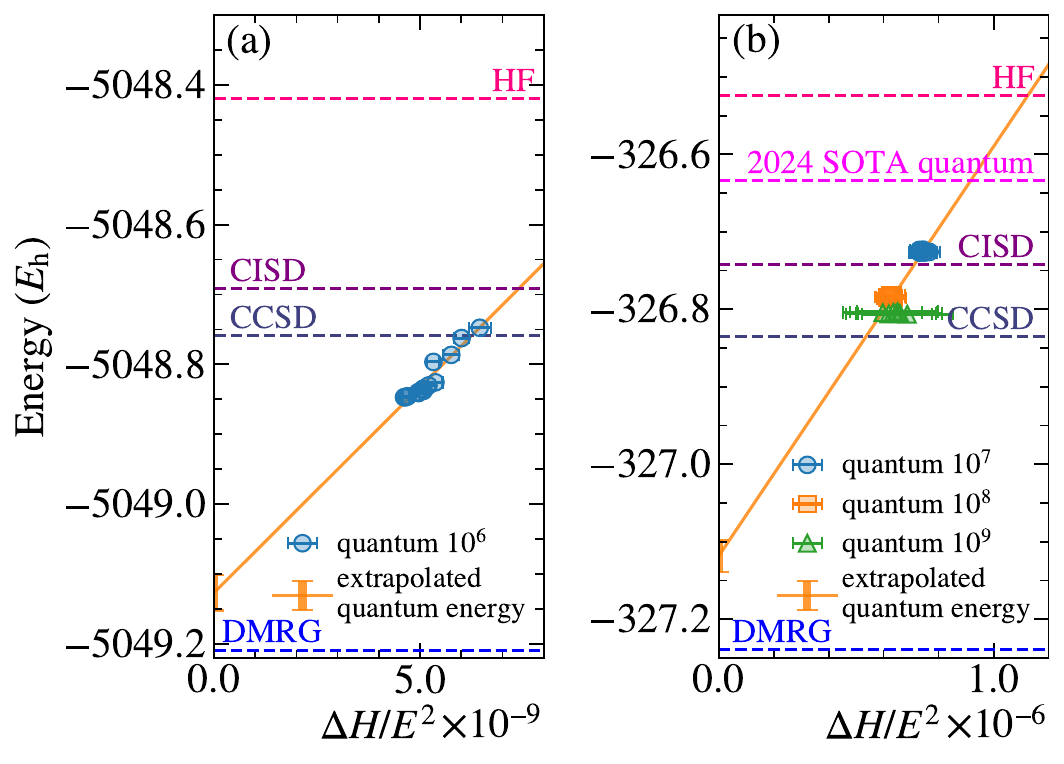}
\caption{{\bf Extrapolated quantum energies}
Quantum SQD energies and corresponding energy variances $\Delta H$ obtained during the last iterations for each subspace dimension for {\bf (a)} [2Fe--2S] and {\bf (b)} [4Fe--4S].
The orange solid line represents a linear fit to the zero-variance quantum energy. The horizontal dashed lines denote, from top to bottom, the Hartree--Fock (HF), configuration interaction with single and double excitations (CISD), coupled--cluster with single and double excitations (CCSD), and density matrix renormalization group (DMRG) energies.
In panel (b), the dashed line corresponding to the 2024 SOTA quantum~\cite{robledo2024chemistry} indicates the best computational quantum result achieved at that time.}
\label{fig:extrap_energy}
\end{figure}

Figure~\ref{fig:extrap_energy} shows the relationship between the energy variance, defined as 
\(\Delta E = \langle \psi \vert \hat{H}^2 \vert \psi \rangle - \langle \psi \vert \hat{H} \vert \psi \rangle^2\),
and the total energy obtained after the optimization loop for the [2Fe–2S] and [4Fe–4S] clusters.
As the dimension of the subspace increases, both the magnitude of the energy variance decreases and the corresponding energy improves. Similar to the results in~\cite{robledo2024chemistry}, we observe linear arrangements of points in the energy-variance plane, which allow for extrapolations to zero variance. 
The extrapolated value toward zero energy variance is approximately \(-327.118 \pm 0.02~\hartree\) for [4Fe–4S],
which deviates by about \(0.12~\hartree\) from the DMRG reference energy of \(-327.239~\hartree\).
Similarly, for the [2Fe–2S] cluster, the extrapolated energy is \(-5049.127 \pm 0.02~\hartree\),
with a deviation of only about \(0.08~\hartree\) from the DMRG reference value of \(-5049.217~\hartree\).
These extrapolated energies, near to the DMRG values, indicate that the improved closed-loop workflow presented here, coupled with massive deployment of classical capabilities, can extract meaningful signal from current generation of quantum devices. This signal can capture a significant amount of quantum correlation in challenging chemistry problems.

\section{Conclusions}

Quantum computers are poised to revolutionize the way we approach a few selected computational tasks: the simulation of quantum matter, specifically the computation of electronic structures for molecular systems, is one of them. In order to extract value from quantum computers, the quantum technology must interact with existing classical HPC facilities, within quantum-centric supercomputing environments~\cite{alexeev2023quantum}. 
For real-world computational tasks that can leverage quantum computing, the challenge ahead is to design workflows that exploit both quantum and classical processors in an optimal way. 

The closed-loop workflow presented here achieves this by using a pre-fault-tolerant {\it Heron} quantum processor, coupled to up to 152,064 nodes of the supercomputer {\it Fugaku}. 
We have shown the capability of these two computing accelerators by addressing electronic structure problems which are beyond the reach of classical exact diagonalization methods: two 36-orbital active spaces for the iron sulfide clusters [2Fe-2S] and [4Fe-4S]. We have improved on the state-of-the-art calculations from quantum data for [4Fe-4S]~\cite{robledo2024chemistry} by $278~m\hartree$, and have shown energy-variance extrapolated energies for both [2Fe-2S] and [4Fe-4S] with an energy error of about 0.1~$\hartree$ from classical state-of-the-art DMRG calculations. 
Although showcased for the iron sulfides, the methods presented are widely applicable across electronic structure problems in quantum chemistry. 

We have used an improved workflow, with a quantum-classical feedback loop and the carryover of configurations between optimization iterations. We have orchestrated quantum and classical resources to minimize idle times and maximize performance, adopting a strategy that varied the number of nodes of {\it Fugaku} used as a function of the quantum runtime, the classical subspace size and the quality of the solution. Although the diagonalization code~\cite{sqd-diag} was specifically designed for Fugaku, we have found that it is well suited for current cloud computing environments and other high-performance computing clusters such as IBM Vela and BlueVela systems~\cite{seelamvela}.

Our results show what can be achieved with quantum data deploying some of the largest HPC resources available today, using the current generation of pre-fault-tolerant {\it Heron} quantum computers. This is important both to understand how much value can be extracted from noisy quantum signals, and how efficiently quantum processors can interact with classical HPC infrastructures. 
Looking forward, it is important to observe that quality of the quantum signal improves exponentially with better quantum error rates, while the overall amount of classical resources  decreases correspondingly, at fixed accuracy. This in turn implies that better solutions can be achieved using less classical resources than what we used in this work, offering a path forward to leverage the quantum technology to reach quantum advantage for realistic problems in computational chemistry.

\section*{Acknowledgments}
Part of this work is funded by project JPNP20017, commissioned by the New Energy and Industrial Technology Development Organization (NEDO).
This work used computational resources of Fugaku provided by RIKEN Center for Computational Science (Project ID: ra010014), and through the HPCI System Research Project (Project ID: hp240496, hp240552, hp240553, hp240554). 
This study is supported by JSPS KAKENHI Grants No. JP21H04446 and No. JP22K03479.
We further acknowledge funding from JST COI-NEXT (Grant No. JPMJPF2221) and the Program for Promoting Research of the Supercomputer Fugaku (Grant No. MXP1020230411) from MEXT,
Japan. Additionally, we appreciate the support provided by the UTokyo Quantum Initiative and the RIKEN TRIP initiative (RIKEN Quantum).

\begin{table*}[t]
  \centering
  \caption{Software versions used in the workflow.}
  \label{tab:software}
  \begin{tabular}{cc}
    \hline
    Software & Version  \\
    \hline
    Python  &  3.11.6     \\
    Qiskit  &   1.4.2  \\
    qiskit-ibm-runtime  & 0.37.0   \\
    ffsim  & 0.0.52   \\
    Qiskit addon of SQD  & 0.10.0  \\
    \hline
  \end{tabular}
\end{table*}


\bibliographystyle{IEEEtran}
\bibliography{main}

\end{document}